\documentclass{svproc}
\usepackage{url,graphicx,multicol}
\usepackage[bottom]{footmisc}
\usepackage{xspace}

\def\darklight{{\it Darklight} }

\def\hmpc{\,h^{-1}{\rm Mpc}}
\def\mhmpc{\,h^{-1}{\rm Mpc}}

\def\P{\mathcal{P}}

\def\vr{{\bf r}}

\def\vpa{{v_\|}}

\usepackage{color}

\begin{document}
\mainmatter              
\title{Measuring the Universe with galaxy redshift surveys
}
\titlerunning{Cosmology with redshift surveys}  
%
\author{L.~Guzzo\inst{1,2} 
\and J. Bel\inst{3} 
\and D.~Bianchi\inst{4}
\and C.~Carbone\inst{1,2},\\
B.R.~Granett\inst{2,1}
\and A.J.~Hawken\inst{2,1,3}
\and F.G. Mohammad\inst{1,2},\\ 
A. Pezzotta\inst{2,1,5} 
\and S.~Rota\inst{6} 
\and M.~Zennaro\inst{1,2} 
}
\authorrunning{Luigi Guzzo et al.} 
%
%
\institute{Dipartimento di Fisica `Aldo Pontremoli', Universit\`a degli Studi di Milano,
Italy
\and
INAF - Osservatorio Astronomico di Brera, 
Milano, Italy
\and
Aix--Marseille Universit\'e, Marseille, France
\and
Institute for Cosmology and Gravitation, University of Portsmouth, UK
\and 
Institute of Space Sciences (IEEC-CSIC), Barcelona, Spain
\and 
INAF - IASF Milano, Italy
}

\maketitle              

\begin{abstract}
  Galaxy redshift surveys are one of the pillars of the current
  standard cosmological model and remain a key tool
  in the experimental effort to understand the origin of
  cosmic acceleration. 
To this end, the
  next generation of surveys aim at achieving sub-percent
  precision in the measurement of the equation of state of dark energy
  $w(z)$ and the growth rate of structure $f(z)$.  This however
  requires comparable control over systematic 
  errors, stressing the need for improved modelling methods.  In this 
contribution 
we review at the introductory level some highlights of the work done in this direction by the
  \darklight project\footnote{\tt http://darklight.fisica.unimi.it}. Supported by an ERC Advanced Grant,
\darklight developed novel techniques for clustering analysis,  
which were tested through numerical
  simulations before being finally applied to galaxy data as in particular those of the 
recently completed VIPERS redshift survey.  We
  focus in particular on: (a) advances on estimating the growth rate
  of structure from redshift-space distortions; (b) parameter
  estimation through global Bayesian reconstruction of the density field
  from survey data; (c) impact of massive neutrinos on large-scale structure measurements. 
 Overall, \darklight has contributed to paving the way for
  forthcoming high-precision experiments, such as {\it
    Euclid}, the next ESA cosmological mission.\footnote{Review
    to appear in {\it Towards a Science Campus in Milan:
A snapshot of current research at Physics Department 'Aldo Pontremoli'
} (2018, Springer, Berlin, in press)}

\end{abstract}
\section{Introduction}
\begin{figure}
\centering
\includegraphics[scale=0.45]{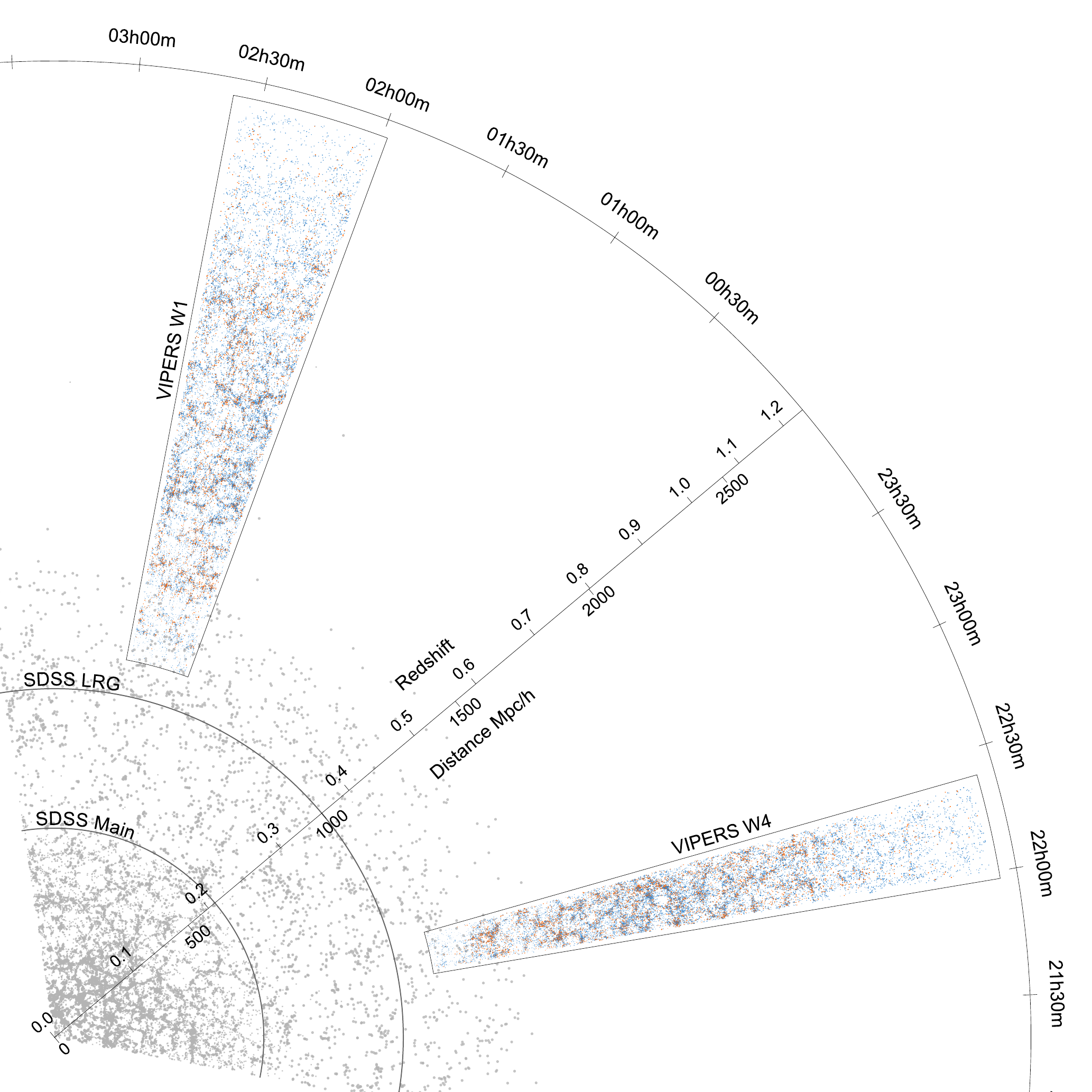}
\caption{Combined ``cone diagram'' of the large-scale distribution of
  galaxies from different surveys, out to
  $z=1$.  The plots includes the recently completed, deep VIPERS survey \cite{guzzo14,garilli14,scodeggio17} and
  two sub-samples of the Sloan Digital Sky Survey (SDSS) (main
  sample and Luminous Red Galaxy (LRG) sample) at lower redshift \cite{york00,eisenstein05}. The
  plotted slices here are 4 and 2 degree-thick for the SDSS and VIPERS
  data, respectively.
}
\label{fig:cones}
\end{figure}

A major achievement in
cosmology over the 20th century has been the detailed reconstruction of
the large-scale structure of the Universe around us.  Started in the
1970s, these studies developed over the following decades into the industry of
{\it redshift surveys}, beautifully exemplified by the Sloan
Digital Sky Survey (SDSS) in its various incarnations (e.g. \cite{eisenstein11}).  These maps have covered in detail our ``local''
Universe (i.e. redshifts $z<0.2$) and only recently we started
exploring comparable volumes at larger redshifts, where
the evolution of galaxies and structure over time can be detected
(see e.g. \cite{guzzo_messenger17}). Fig.~\ref{fig:cones} shows
a montage using data from some of these surveys, providing a visual impression
of the now well-established sponge-like topology of the large-scale galaxy distribution
and how it stretches back into the younger Universe.

In addition to their purely cartographic beauty, these maps provide a quantitative
test of the theories of structure formation and of the Universe composition.  Statistical measurements
of the observed galaxy distribution represent in fact one of the
experimental pillars upon which the
current ``standard'' model of cosmology is built.  
Let us define the matter over-density (or fluctuation) field, 
with respect to the mean density, as
$\delta(\mathbf{x}) \equiv (\rho(\mathbf{x}) - \bar{\rho})/\bar{\rho}$; this can be described in terms 
of Fourier harmonic components as 
 \begin{equation}
 \delta(\mathbf{k})=  \int_V \delta(\mathbf{x}) \, e^{-i
     \mathbf{k} \cdot \mathbf{x}} \, d^3 \mathbf{x}   \, ,
 \end{equation}
 where $V$ is the volume considered.  The power spectrum $P(\mathbf{k})$ is
 then defined by the variance of the Fourier modes: 
 \begin{equation}
 \langle \delta(\mathbf{k}) \delta^*(\mathbf{k'}) \rangle = (2\pi)^3
 P(\mathbf{k}) \delta_{\rm D}(\mathbf{k} - \mathbf{k'}) \,\,\, . 
 \end{equation}
 The observed number density of galaxies $n_g(\mathbf{x})$ is related to the matter fluctuation field through the {\it bias parameter} $b$ by
\begin{equation}
n_g = \bar{n} \left(1 + b \delta \right), 
\label{eq:bias}
\end{equation}
which corresponds to assuming that $\delta_{g}=
b\delta$.  
This linear and scale-independent relation provides 
an accurate description of galaxy clustering at large scales, although 
it breaks down in the quasi-linear regime below scales of 
$\sim10\mhmpc$ \cite{diporto16}. In general, $b$ depends on 
galaxy properties, as we shall discuss in more detail in Sect.~\ref{sec:bayes}.
From the hypothesis of linear bias, it descends that $P_{gg}(k)=b^2 P(k)$, where 
$P_{gg}(k)$ is the observed galaxy-galaxy power spectrum. This 
connection allows us to use measurements of $P_{gg}(k)$ to  
constrain the values of cosmological parameters that regulate the shape of $P(k)$.
%
 \begin{figure}
 \vspace{-0.4cm}
 \centering
 \includegraphics[scale=0.26]{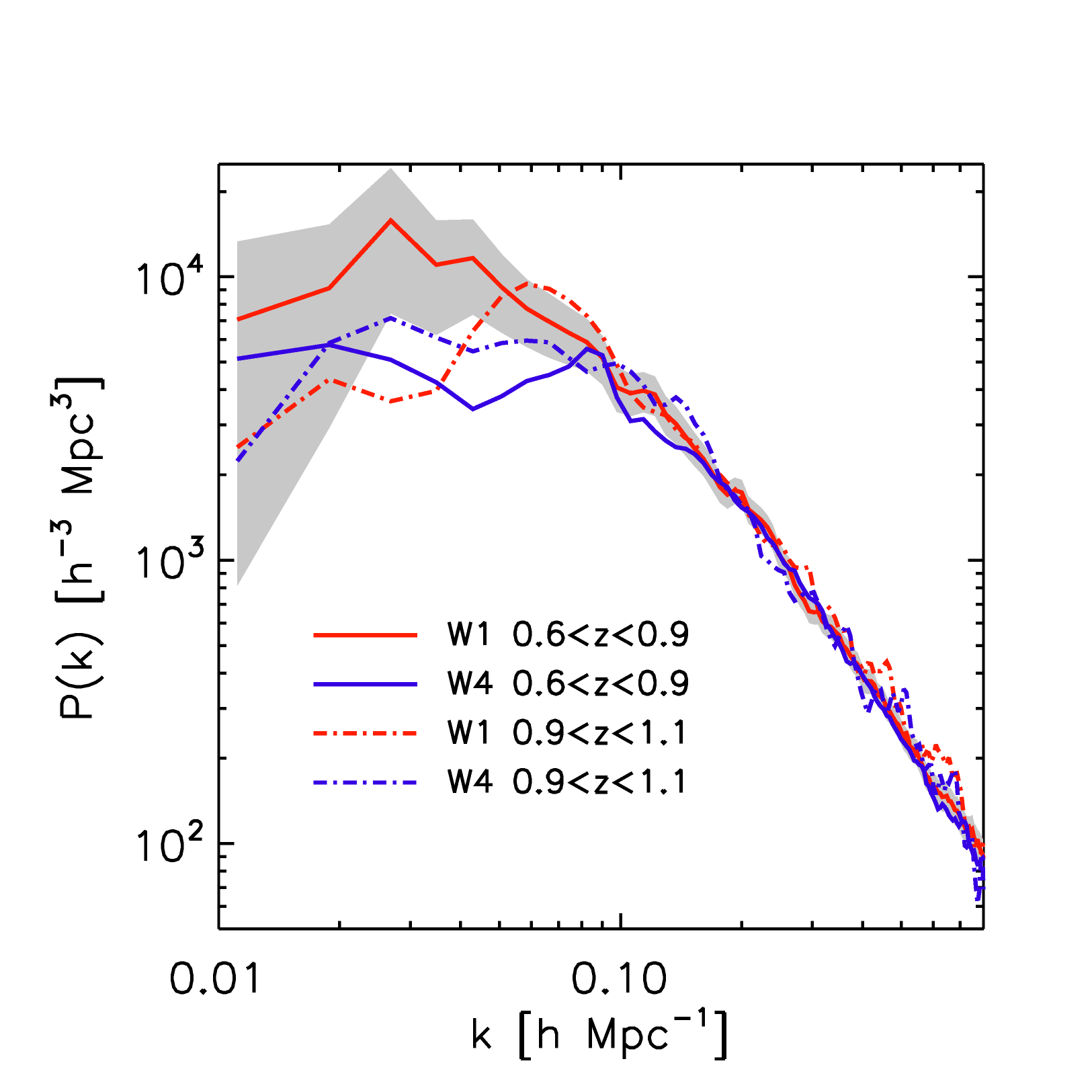}\hspace{-0.4cm}
 \includegraphics[scale=0.245]{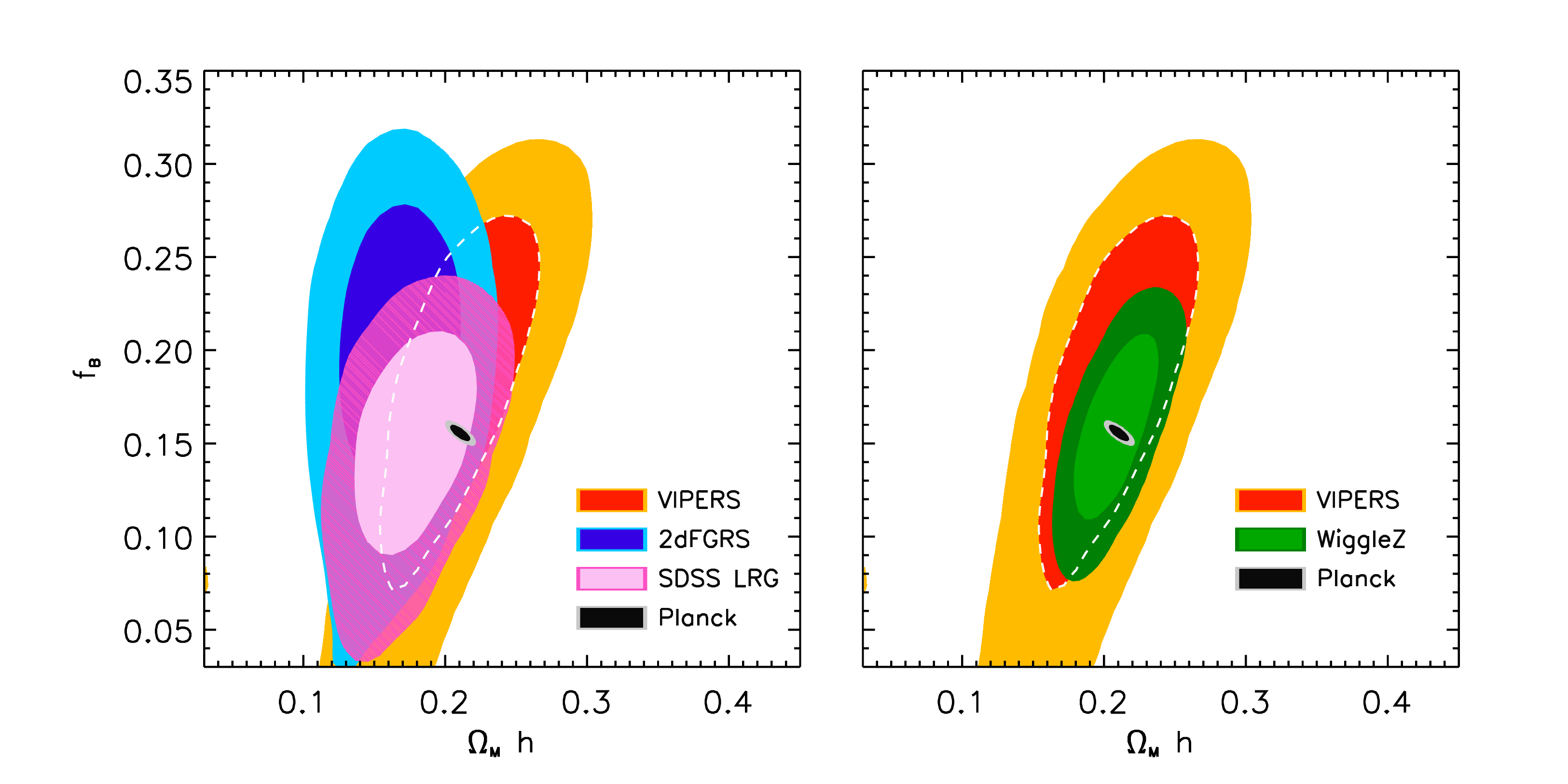}
 \caption{{\it Left}: Four independent estimates of the power spectrum of the galaxy
    distribution at $0.6<z<1.1$ from the VIPERS galaxy survey. The four curves correspond to two redshift
    bins for the two separated fields W1 and W4, which have slightly
    different window functions (i.e. size and geometry). {\it Center \& Right:}
    VIPERS constraints on the
  mean total matter density $\Omega_M$ times the normalised Hubble
  constant $h=H_o/100$ and the baryonic fraction, 
  $f_b=\Omega_b/\Omega_M$, compared to similar measurements from surveys at low 
  (center) and high redshift (right), plus Planck.
 See \cite{rota17} for details.}
 \label{fig:pk}
 \end{figure}
 Fig.~\ref{fig:pk} \cite{rota17} shows an example of such measurements: the left panel plots four estimates 
 of the power spectrum $P(k)$ (more precisely, its monopole, i.e. the average of
 $P(\mathbf{k})$ over spherical shells) obtained at $0.6<z<1.1$ from the VIPERS survey data of 
 Fig.~\ref{fig:cones} (see also Sect.~\ref{sec:data}). In the central and right panels, we show the posterior distribution 
 of the mean density of matter $\Omega_m$ and the baryon fraction $f_B$ from
 a combined likelihood analysis of the four measurements; these are compared to
 similar estimates from other surveys and from the Planck CMB anisotropy constraints \cite{planck15}.
 More precisely, the galaxy power spectrum shape on large scales probes the combination $\Omega_Mh$, where $h=H_o/100$.
 Such comparisons provide us with important tests of the $\Lambda$CDM model, with the $z\sim 1$ estimate
 from VIPERS straddling Planck and local measurements. 

If one goes beyond the simple shape of angle-averaged quantities, 
two-point statistics of the galaxy distribution contain further powerful information,
which is key to understanding the origin of the mysterious acceleration of cosmic
expansion discovered less than twenty years ago \cite{riess98,perlmutter99}.  
First, tiny ``baryonic wiggles'' in the shape
of the power spectrum define a specific, well known comoving spatial
scale, corresponding to the sound horizon scale at the epoch when
baryons were dragged into the pre-existing dark-matter potential
wells.  In fact, it turns out that there
are enough baryons in the cosmic mixture to influence the
dominant dark-matter fluctuations \cite{cole05,eisenstein05} and leave in the
galaxy distribution a visible signature of the pre-recombination acoustic oscillations
in the baryon-radiation plasma.  Known as Baryonic Acoustic
Oscillations (BAO), these features provide us with a formidable standard ruler to measure the expansion
history of the Universe $H(z)$, complementary to what can be done 
using Type Ia supernovae as standard candles (see e.g. \cite{alam17} for the latest measurements from the SDSS-BOSS sample).

Secondly, the observed redshift maps are distorted by the contribution of peculiar
velocities that cannot be separated from the cosmological
redshift. This introduces a measurable anisotropy in our clustering
statistics, what we call Redshift Space Distortions (RSD), an effect that provides us
with a powerful way to probe the {\it growth rate of 
structure} $f$.  This key information can break the degeneracy on whether
the observed expansion history is due to the presence of the extra
contribution of a cosmological constant (or
dark energy) in Einstein's equations or rather require a more 
radical modification of gravity theory.  While RSD were first described in the 1980's \cite{davis83,kaiser87}), their 
potential in the context of understanding the origin of cosmic acceleration
was fully recognized only recently \cite{guzzo08}; nowadays they are considered one of the potentially most powerful ``dark energy tests'' expected from the next generation of cosmological surveys, as in particular the ESA mission
{\it Euclid} \cite{laureijs11}, of which the Milan group is one of the original founders.
\section{Measuring the growth rate of structure from RSD}
\subsection{Improved models of redshift-space distortions}
\label{sec:mod}
Translating galaxy clustering observations into precise and accurate 
measurements of the key cosmological parameters, however, requires modelling the effects of
non-linear evolution, galaxy bias (i.e. how galaxies trace mass) and
redshift-space distortions themselves.  The interest in RSD
precision measurements 
stimulated work to verify the accuracy of these measurements \cite{okumura11,bianchi12}. Early estimates -- focused essentially on measuring $\Omega_M$, given that in the context of General Relativity $f\simeq \Omega_M^{0.55}$ (e.g. \cite{peacock01}) -- adopted empirical non-linear corrections to the original linear theory by Kaiser; this is the case of the so-called ``dispersion model'' \cite{peacock94}, which in terms of the power spectrum of density fluctuations is expressed as
\begin{equation}
P^s(k,\mu)=D\big(k\mu\sigma_{12}\big)\,\Bigg(1+\beta\mu^2\Bigg)^2b^2P_{\delta\delta}(k),
\label{eq:disp}
\end{equation}
where $P^s(k,\mu)$ is the redshift-space power spectrum, which depends
both on the amplitude $k$ and the orientation $\mu=\cos(\Phi)$ of
the Fourier mode with respect to the line-of-sight, 
$P_{\delta\delta}(k)$ is the real-space (isotropic) power spectrum of
the matter fluctuation field $\delta$ and $\beta=f/b$, with $f$ being the growth of structure and $b$  the {\it linear bias} of the
specific population of halos (or galaxies) used. The latter is defined as the ratio of the {\it rms} clustering amplitude of galaxies to that of the matter, conventionally measured in spheres of $8 \hmpc$ radius, $b=\sigma_8^{gal}/\sigma_8$. For what will follow later, it is useful to note that
\begin{equation}
\beta=\frac{f}{b}=f\frac{\sigma_8}{\sigma_8^{gal}} ,
\end{equation}
can be recast as
\begin{equation}
\beta \sigma_8^{gal} = {f}{\sigma_8} ,
\end{equation}
which combines two directly measurable quantities to the left, showing that what we actually measure is the combination of the growth rate and the {\it rms} amplitude of clustering, ${f}{\sigma_8}$. This is what nowadays is customarily plotted when presenting measurements of the growth rate from redshift surveys (e.g. Fig.~\ref{fig:fsig8}).

\begin{figure}
\centering
\includegraphics[scale=0.7]{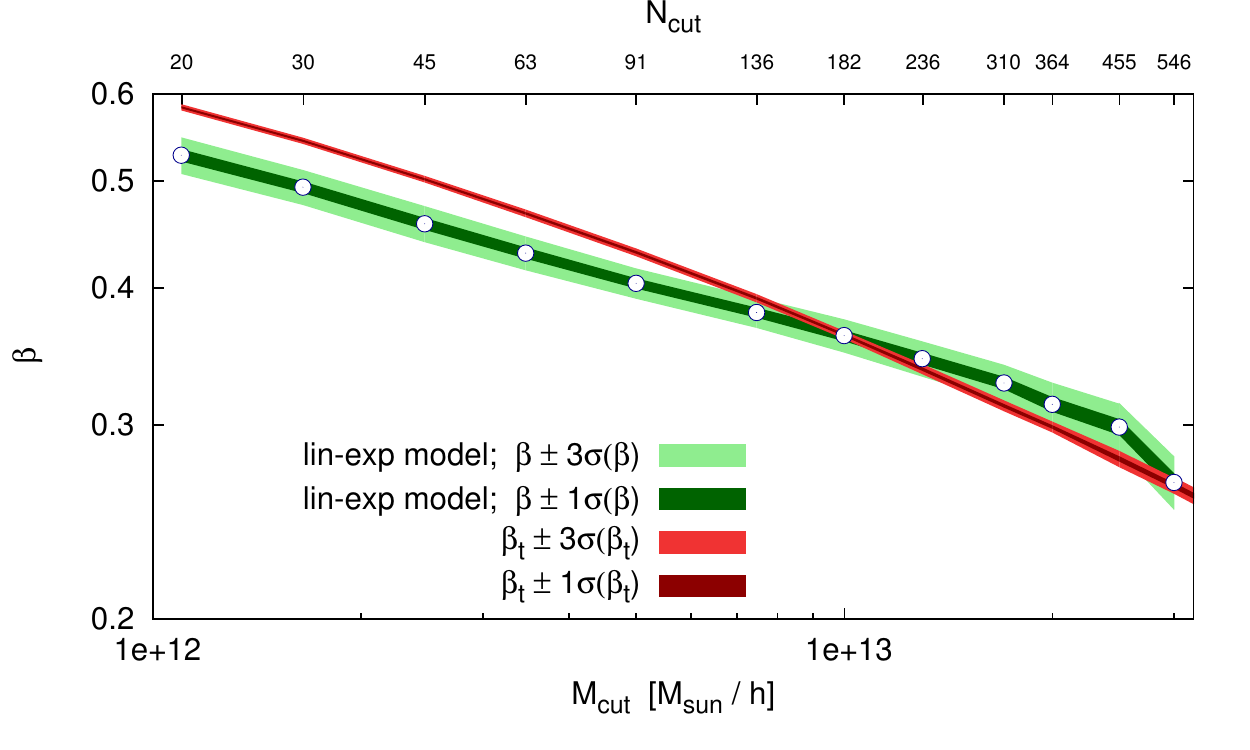}
\caption{Systematic differences in the measured values of the RSD distortion parameter
  $\beta$ (dots with green 1- and 2-$\sigma$ error bands) with respect
  to the expected value (thinner red band, including theoretical
  uncertainties). Measurements are performed for catalogues of dark-matter halos with increasing threshold mass, built from an n-body simulation 
\cite{bianchi12}}
\label{fig:syst1}
\end{figure}

Going back to eq.~(\ref{eq:disp}), the term
$D\big(k\mu\sigma_{12}\big)$ is usually either a Lorentzian or a Gaussian function, empirically introducing a nonlinear damping to the Kaiser linear amplification, with the Lorentzian (corresponding to an exponential in configuration space) normally providing a better fit to the galaxy data \cite{pezzotta17}. This term is regulated by a second
free parameter, $\sigma_{12}$, which corresponds to an effective (scale-independent) line-of-sight pairwise velocity dispersion.  Fig.~\ref{fig:syst1} (from \cite{bianchi12}), shows how estimates of $\beta$ using the dispersion model can be plagued by systematic errors as large as 10\%, depending on the kind of galaxies (here dark matter halos) used.  With the next generation of surveys aiming at 1\% precision by collecting several tens of millions of redshifts, such a level of systematic errors is clearly unacceptable. 

Exploring how to achieve this overall goal by optimising measurements of galaxy clustering and RSD, has been one of the main goals of the
\darklight project, supported by an ERC Advanced Grant awarded in 2012.  \darklight focused on developing new techniques, testing them on simulated samples, and then applying them to the new data from 
the VIMOS Public Extragalactic Redshift Survey (VIPERS), which was built in parallel.
%

After assessing the limitations of existing RSD
models \cite{bianchi12,delatorre12}  the first goal of \darklight has
been to develop refined theoretical descriptions. This work followed two branches: one,
starting from first principles, was based on revisiting the so-called {\it
  streaming model} approach; the second, more pragmatic, aimed at
refining the application to real data of the best models available at the time, 
as in particular the ``TNS'' model \cite{taruya10}.  Such more
``data oriented'' line of development also included exploring the advantages
of specific tracers of large-structure in reducing the
impact of non-linear effects.

The first approach
\cite{bianchi15} focused on the so-called {\it
  streaming model} \cite{fisher95}, which in the more
general formulation by Scoccimarro \cite{scoccimarro04} (see also \cite{reid12}), describes the two-point
correlation function in redshift space
$\xi_S(s_\bot, s_\|)$ as a function of its real-space counterpart $\xi_R(r)$
\begin{equation}
\label{eq streaming}
1 + \xi_S(s_\bot , s_\|) = \int dr_\| \ [1 + \xi_R(r)] \ \P(r_\| - s_\| | \vr) \ .
\end{equation}
Here quantities noted with $\bot$ and $\|$ correspond to the components of the pair
separation -- in redshift or real space -- respectively perpendicular and parallel to the line of
sight, with $r^2=r_\|^2+r_\bot^2$ and $r_\bot=s_\bot$.
The interest in the streaming model is that this expression is exact: knowing
the form of the pairwise velocity distribution function $\P(\vpa |
\vr) = \P(r_\| - s_\| | \vr)$ at any separation 
$\vr$, a full mapping of real- to redshift-space correlations is
provided.  The problem is that this is a virtually infinite family of
distribution functions.

The essential question addressed in
\cite{bianchi15} has been whether a sufficiently accurate description of
this family (and thus of RSD) is still
possible with a reduced number of degrees of freedom.  It is found
that, at a given galaxy separation $\vr$, they can be described as a superposition
of virtually infinite Gaussian functions, whose mean $\mu$ and dispersion
  $\sigma$ are in turn distributed according to a bivariate Gaussian,
  with its own mean and covariance matrix.  A recent extension of
  this work \cite{bianchi16} shows that such``Gussian-Gaussian'' model cannot
  fully match the level of skewness observed at small separations, in
  particular when applied to catalogues of dark matter halos. 
  They thus generalize the model by allowing for the presence of a small amount of local skewness, meaning that the velocity distribution is obtained as a superposition of quasi-Gaussian functions.
  In its simplest formulation, this improved model takes as input the real space correlation function and the first three velocity moments (plus two well defined nuisance parameters) and returns an accurate description of the anisotropic redshift-space two-point correlation function down to very small scales ($\sim 5 \mhmpc$ for dark matter particles and virtually zero for halos).
  To be applied to real data to estimate the growth rate of structure $f$, the model still needs a better theoretical and/or numerical understanding of how the velocity moments depend on $f$ on small scale, as well as tests on mock catalogues including realistic galaxies. 

The second, parallel approach followed in \darklight was to work on the
``best'' models existing in the literature, optimising their
application to real data.  The natural extensions to the dispersion
model (\ref{eq:disp}) start from the Scoccimarro \cite{scoccimarro04} expression
\begin{equation}
P^s(k,\mu)=D\big(k\mu\sigma_{12}\big)\,\Big(b^2P_{\delta\delta}(k)+2fb\mu^2P_{\delta\theta}(k)+f^2\mu^4P_{\theta\theta}(k)\Big), 
\label{eq:scoccimarro_model}
\end{equation} 
where $P_{\delta\theta}$ and $P_{\theta\theta}$ are respectively the so-called
density-velocity divergence cross-spectrum and the velocity divergence
auto-spectrum, while $P_{\delta\delta}$ is the usual matter power
spectrum.  If one then also accounts for the non-linear mode 
coupling between the density and velocity-divergence fields, two more
terms arise inside the parenthesis, named $C_A(k,\mu,f,b)$ and $C_B(k\,u,f,b)$, leading to the
TNS model by Taruya and collaborators \cite{taruya10}.
%

A practical problem in the application of either of these two models is that the
values of $P_{\delta\theta}$ and $P_{\theta\theta}$  cannot be measured from the
data.  As such, they require
empirical fitting functions to be calibrated using numerical
simulations \cite{jennings11}.  As part of the \darklight work, we used the DEMNUni simulations (see sect.~\ref{sec:neutrinos}) to derive improved fitting functions in different cosmologies \cite{bel17}:
\begin{equation}
P_{\delta\theta}(k)=\bigg(P_{\delta\delta}(k)P^{lin}(k)e^{-k/k^*}\bigg)^{\frac{1}{2}},
\label{eq:fit_pdt}
\end{equation}
\begin{equation}
P_{\theta\theta}(k)=P^{lin}(k)e^{-k/k^*},
\label{eq:fit_ptt_1p}
\end{equation}
where $P^{lin}(k)$ is the linear matter power spectrum and $k^*$ is a
parameter representing the typical damping scale of the velocity power
spectra, which is well described as $1/k^*=p_1\sigma_8^{p_2}$,
where $p_1,\,p_2$ are the only two parameters that need to be calibrated from the simulations.
These forms for $P_{\delta\theta}$ and $P_{\theta\theta}$ have
valuable, physically motivated properties: they naturally converge to
$P_{\delta\delta}(k)$ in the linear regime, including a dependence on
redshift through $\sigma_8(z)$.  They represent a significant improvement over
previous implementations of the Scoccimarro and TNS 
models and allowed us to extend their application to smaller scales
and to the high redshifts covered by VIPERS.

\subsection{Application to real data: optimising the samples}
\label{sec:data}
%
The performance, in terms of systematic error, of any RSD model when 
applied to real data does not depend only on the quality of the model itself. 
The kind of tracers of the density and velocity field that are used, 
significantly enhance or reduce some of the effects 
we are trying to model and correct. This means that, in principle, we may be
able to identify specific sub-samples of galaxies for which the needed 
non-linear corrections to RSD models are intrinsically smaller. This could be
an alternative to making our models more and more complex, as it happens for 
the full galaxy population. 
\begin{figure}
\vspace{-0.5cm}
\centering
\includegraphics[scale=0.5]{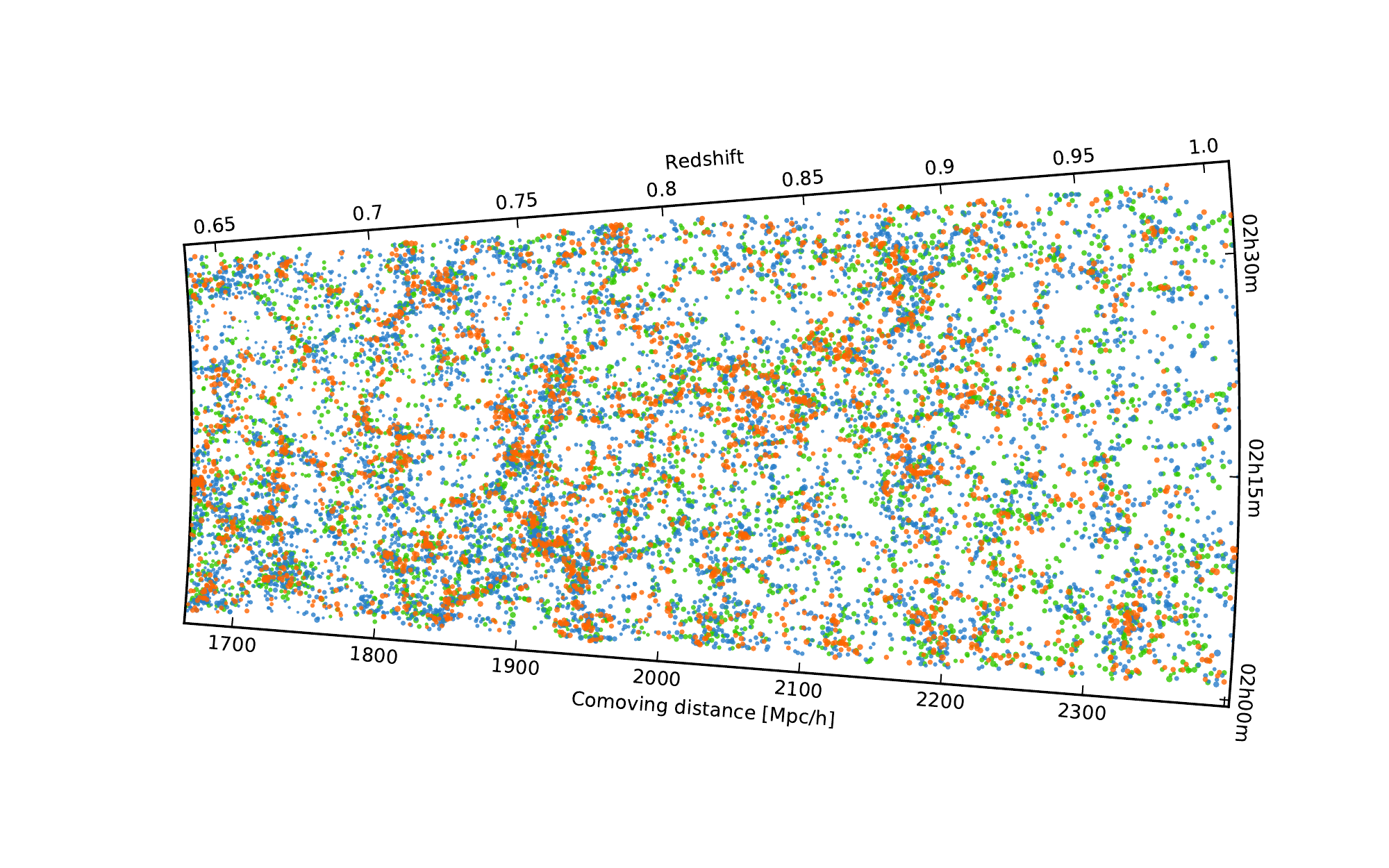}
\caption{A zoom into the central part of the W1 VIPERS
  region. Galaxies are described by dots, whose size is proportional
  to the $B$-band luminosity of the galaxy and whose colour
  corresponds to the its actual $(U–B)$ restframe colour. Note the
  clear colour–density relation, for the first time seen so clearly at
  these redshifts, with red early-type galaxies tracing the backbone
  of structure and blue/green star-forming objects filling the more
  peripheral lower-density regions.}  
\label{fig:zoom}
\end{figure}

Such an approach becomes
feasible if the available galaxy survey was constructed with a broad
selection function and supplemented by extensive ancillary information
(e.g. multi-band photometry, from which spectral energy distributions, 
colours, stellar masses, etc. can be obtained). This allows a wide space 
in galaxy physical properties to be explored, experimenting with clustering
and RSD measurements using different classes of tracers (and their combination),
as e.g. red vs. blue galaxies, groups, clusters. This is the case, for example, of 
the Sloan Digital Sky Survey main sample \cite{york00}.
The VIMOS Public Extragalactic Redshift
Survey (VIPERS) \cite{guzzo14} was designed with the idea of extending this concept to $z\sim 1$,
i.e. when the Universe was around half its current age, providing \darklight with
a state-of-the-art playground.

VIPERS is a new statistically complete redshift survey,
constructed between 2008 and 2016 as one of the ``ESO Large Programmes", exploiting the 
unique capabilities of the VIMOS multi-object spectrograph at the
Very Large Telescope (VLT) \cite{scodeggio17}.  It has secured redshifts for $86,775$ galaxies 
with magnitude $i_{AB} \le 22.5$ (out of $
97,714$ spectra) over a total area of $23.6$ square degrees, 
tiled with a mosaic of 288 VIMOS pointings.  Target galaxies
were selected from the two fields (W1 and W4) of the 
Canada-–France-–Hawaii Telescope Legacy Survey Wide catalogue (CFHTLS--Wide),
benefiting of its excellent image 
quality and photometry in five bands ($ugriz$)\footnote{\tt http://www.cadc-ccda.hia-iha.nrc-cnrc.gc.ca/en/cfht}.
The survey concentrates over the range $0.5<z<1.2$, thanks to a robust colour pre-selection
that excluded lower-$z$ targets, nearly doubling in this way the sampling density achieved 
by VIMOS within the redshift of interest \cite{guzzo14}. This set-up produces 
a combination of dense sampling ($> 40\%$) and large volume ($\sim 5
\times 10^7$ h$^{–-3}$ Mpc$^3$), which is unique for these redshifts and allows studies of
large-scale structure and galaxy evolution to be performed on equal statistical footing with
state-of-the-art surveys of the local $z<0.2$ Universe (see Fig.~\ref{fig:cones}).  
Sparser samples like the SDSS LRG, BOSS \cite{alam17} 
or Wigglez \cite{blake10} surveys allow for much larger volumes to be probed and are
excellent to measure large-scale features as Baryonic Acoustic
Oscillations. However, they include a very specific, limited sample of
the overall galaxy population and (by design) fail to register the details of the underlying nonlinear structure.  
The rich content of information of
VIPERS can be further appreciated in Fig.~\ref{fig:zoom}, where the
connection between galaxy colours and large-scale structure is readily visible by eye.
\begin{figure}
\centering
\includegraphics[scale=0.25]{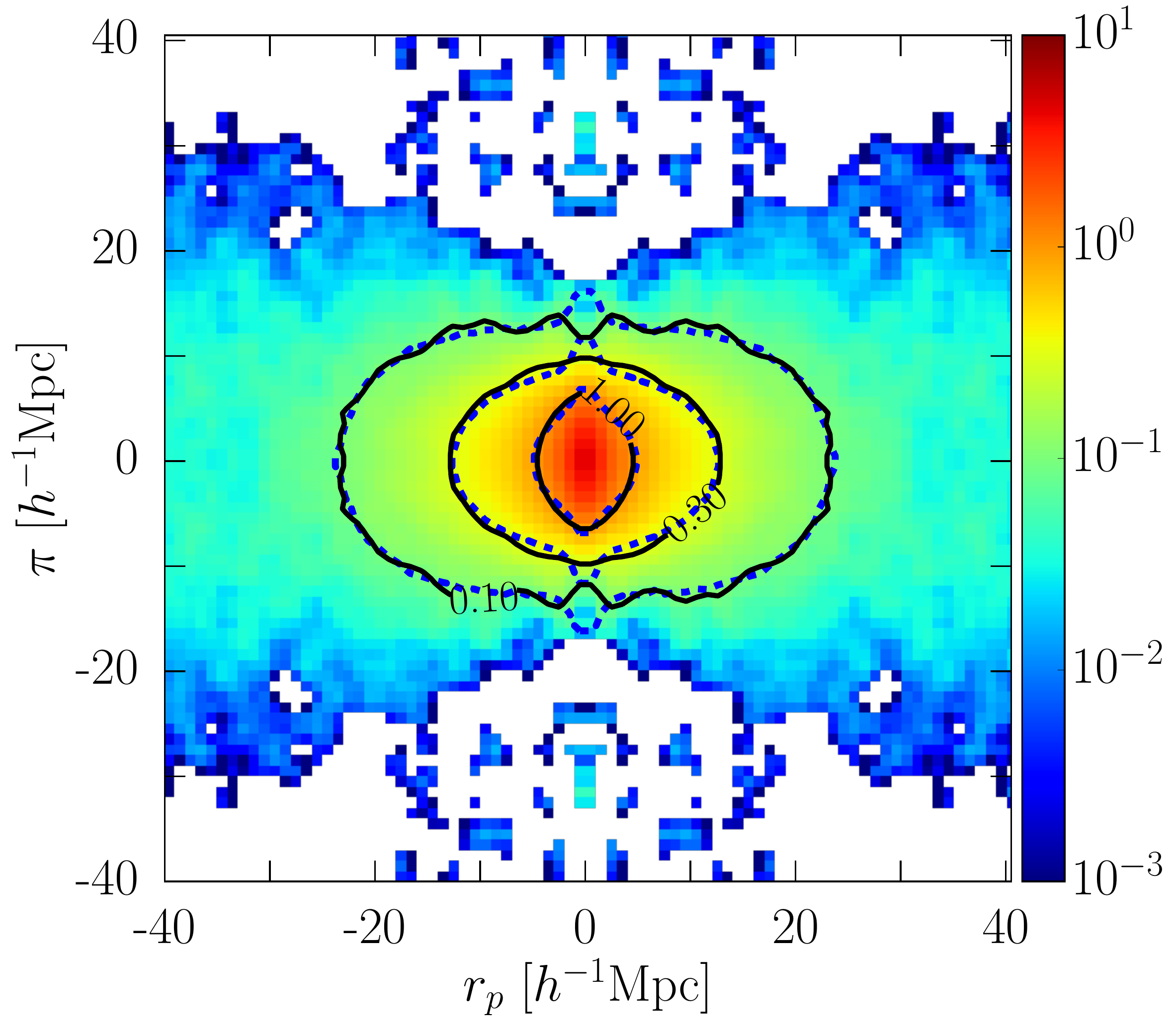}
\includegraphics[scale=0.25]{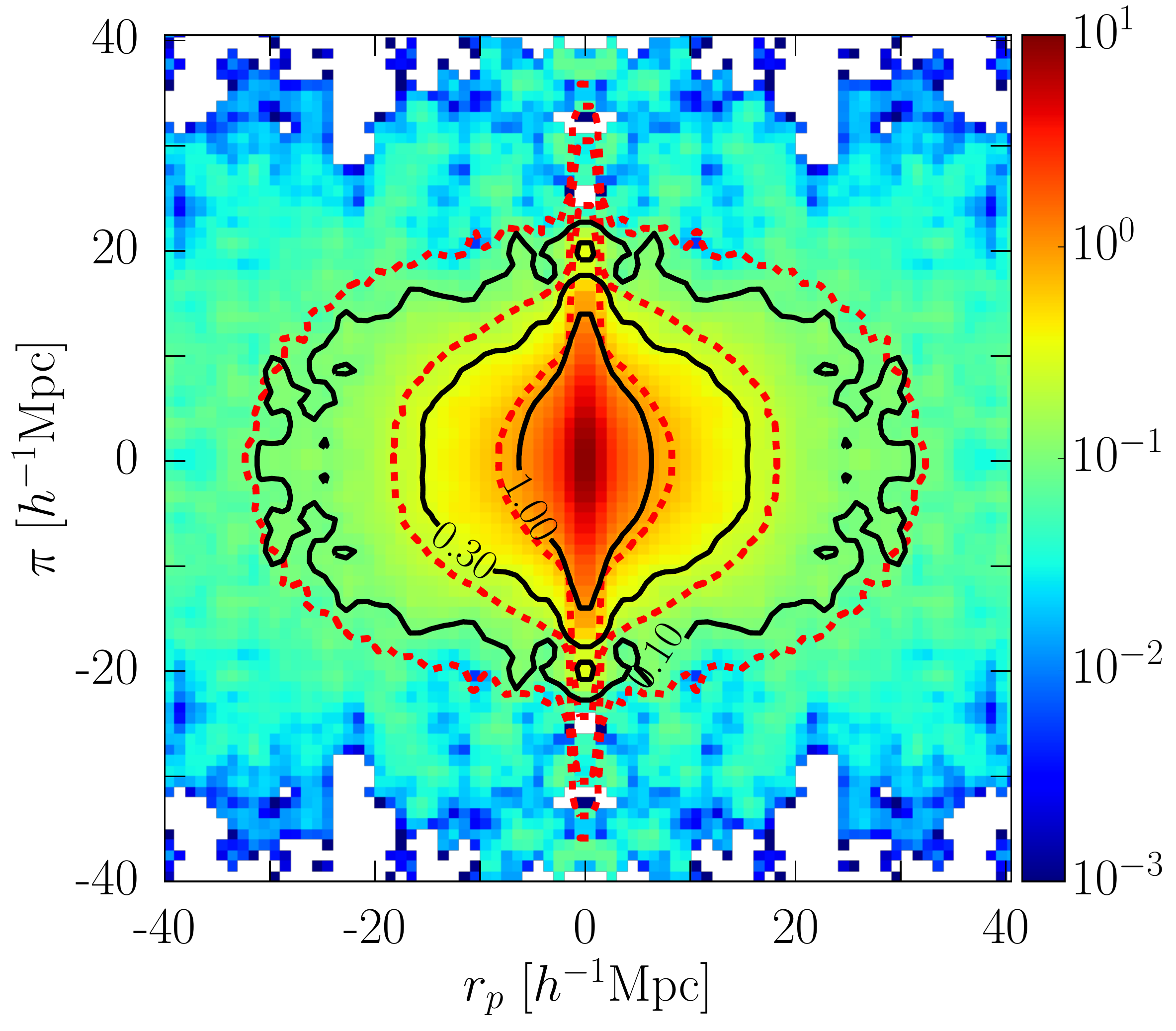}
\caption{Estimate of the redshift-space two-point correlation functions from the VIPERS survey, splitting the sample into blue (left) and red (right) galaxies (colour scale and solid contours),
compared to measurements from a set of mock samples (dashed lines). Blue galaxies show reduced stretching along the line-of-sight ($\pi$) direction, indicating lower contribution by non-streaming velocities, which are the most difficult to account for in the extraction of the linear component and the growth rate of structure $f$ \cite{mohammad17}. } 
\label{fig:xip}
\end{figure}
\begin{figure}
\hspace{-0.2cm}
\includegraphics[scale=0.5]{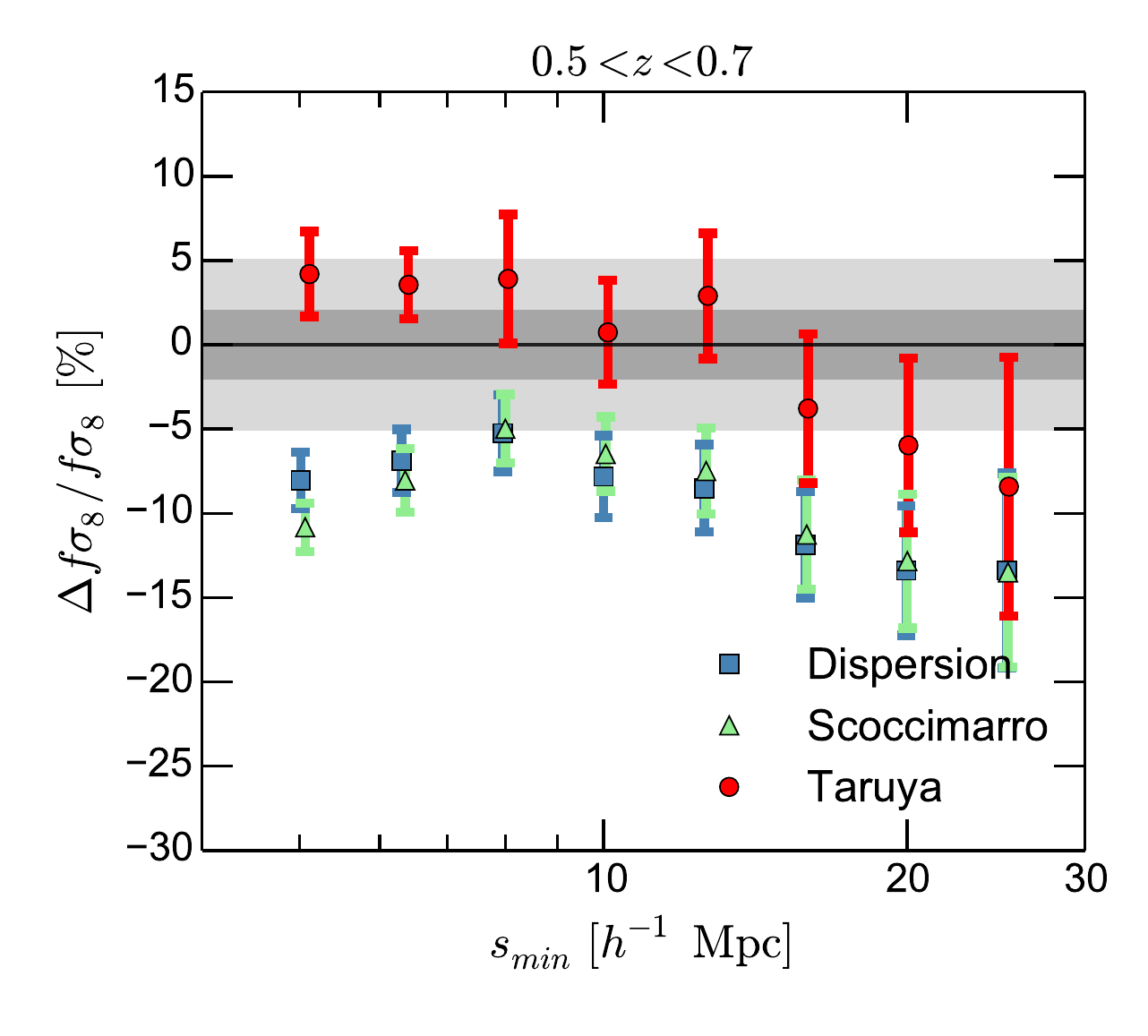}
\includegraphics[scale=0.14]{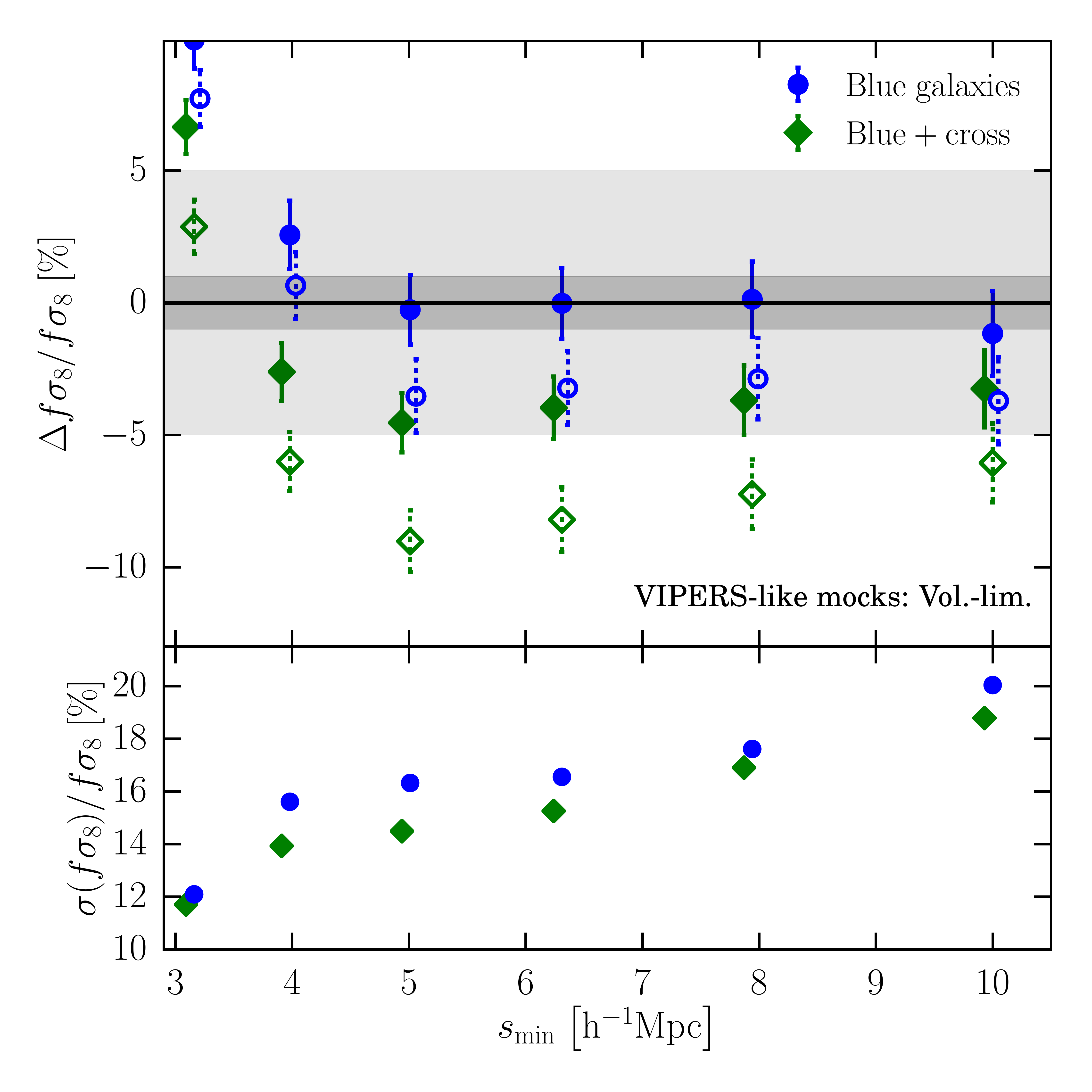}
\caption{Systematic errors on the growth rate parameter
  $f\sigma_8$ using 153 VIPERS-like mock
  catalogues.  In both panels the abscissa correspond to the minimum scale included in
  the fit: the smaller $s_{min}$, the more nonlinear effects are
  included. {\it Left:} improving non-linear corrections in the RSD 
  model \cite{pezzotta17}. {\it Right:} improving the galaxy tracers: 
  luminous blue galaxies yield negligible systematic errors  
  down to $5\hmpc$, even limiting non-linear corrections to the Scoccimarro extension (filled circles)
  of the dispersion model (open circles). } 
\label{fig:syst}
\end{figure}
VIPERS released publicly its final catalogue and a series of new scientific
results in November 2016. More details on the survey
construction and the properties of the sample can be found in
\cite{scodeggio17,garilli14,guzzo14}. 
%

 Fig.~\ref{fig:xip} shows two measurements of  the anisotropic two-point correlation
function in redshift space (i.e. what is called $\xi_S(s_\bot, s_\|)$ in eq.~(\ref{eq streaming}); here $r_p = s_\bot$ and $ \pi = s_\|$), using the VIPERS data. In this case the
sample has been split into two classes, i.e. blue and red galaxies,
defined on the basis of their rest-frame $(U-V)$ photometric colour
(see \cite{mohammad17} for details).   The signature of the linear streaming motions produced by
the growth of structure is evident in the overall flattening of the
contours along the line-of-sight direction ($\pi$).  These plots also
show how blue galaxies (left) are less affected by small-scale
nonlinear motions, i.e. those of high-velocity pairs within virialised
structures.  These produce the small-scale streching of the contours along $\pi$
(vertical direction), which is instead evident in the central part of
the red galaxy plot on the right.   For this reason, blue galaxies turn
out to be better tracers of RSD, for which it is sufficient to use a simpler
modelling, as shown in Fig.~\ref{fig:syst}. When using the full galaxy 
population, the best performing model is the TNS by Taruya et al. \cite{taruya10} 
(left panel), while when we limit the sample
to luminous blue galaxies only, it is sufficient to use the simpler 
nonlinear corrections by Scoccimarro \cite{scoccimarro04} (filled circles, right
panel); open circles correspond to the simplest model, i.e. the standard
  dispersion model \cite{peacock94}, which is not sufficient
  even in this case.  See \cite{mohammad17} for details.

\subsection{RSD from galaxy outflows in cosmic voids}
\begin{figure}
\centering
\includegraphics[scale=0.13]{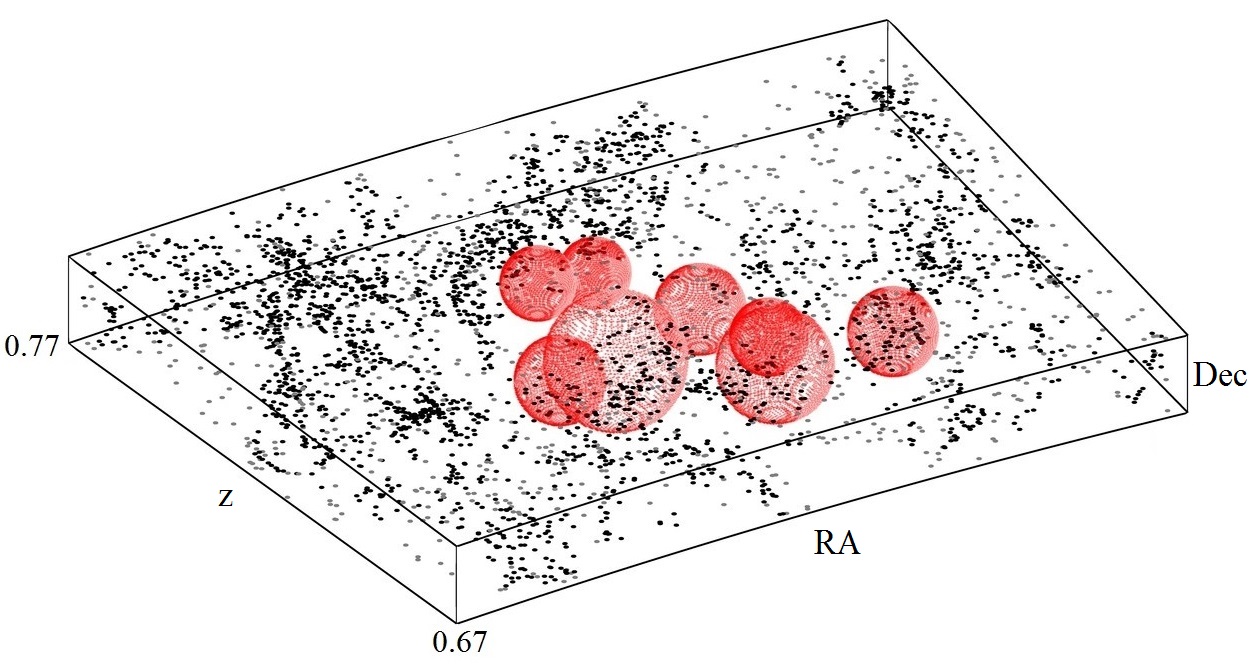}
\includegraphics[scale=0.13]{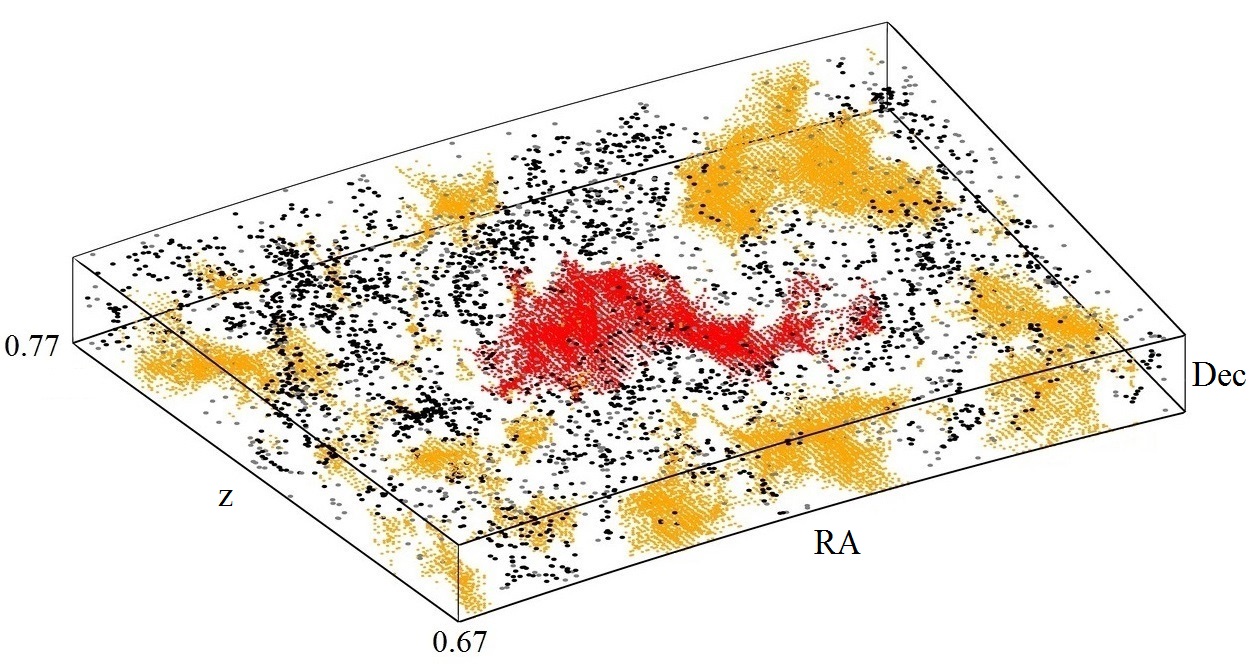}
\caption{Example of definition and search for "voids", as performed in VIPERS. 
{\it Left:} the spherical void regions that make up the largest
  void in one of the VIPERS fields. {\it Right:} in red, the
  centres of all overlapping significant spheres defining the same low-density region; other void regions
  within this volume are shown in orange \cite{micheletti14}.}  
\label{fig:voids}
\end{figure}

Cosmic voids, i.e. the large under-dense regions visible also in Fig.~\ref{fig:cones}, 
represent an interesting new way to look at the data from
galaxy redshift surveys.  As loose as they may appear, over the past
few years they have proved to be able to yield quantitative
cosmological constraints on the growth of structure.
Indeed, growth-induced galaxy peculiar velocities tend to outflow radially from voids,
which leaves a specific mark in the observed void-galaxy
cross-correlation function (see e.g. \cite{hamaus16}). 
The dense sampling of VIPERS makes it excellent for looking for
cosmic voids at high redshift. Fig.~\ref{fig:voids} 
shows an example of how a catalogue of voids was
constructed from these data \cite{micheletti14}.  

\begin{figure}
\centering
\includegraphics[scale=0.23]{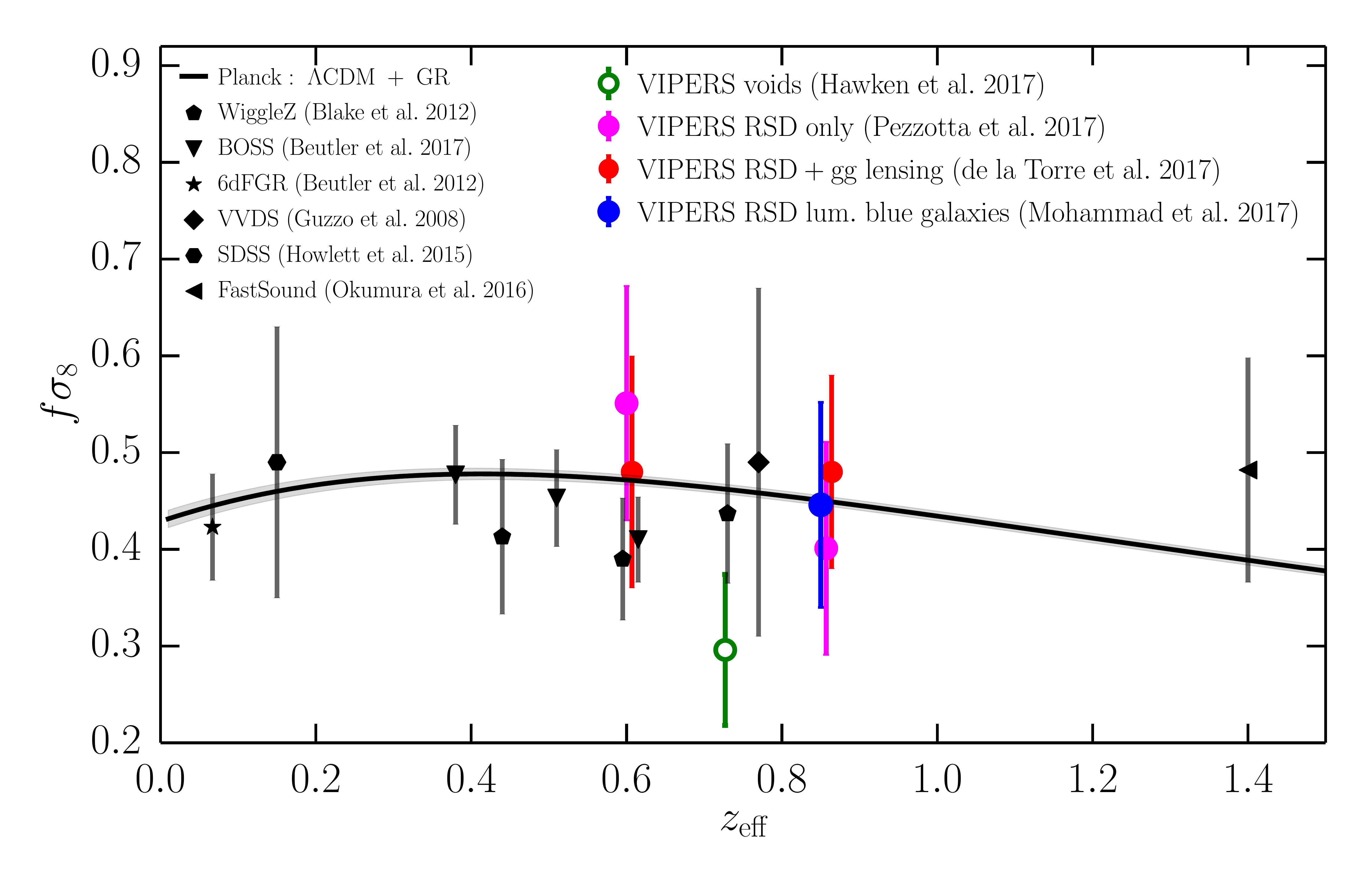}
\caption{Collection of all VIPERS estimates of the growth rate of structure
using complementary methods, compared to literature results 
\cite{blake12,beutler17,beutler12,guzzo08,howlett15,okumura16}, and the prediction of 
the standard cosmological model with current Planck parameter 
values and uncertainties (solid black/grey band) \cite{planck15}.  
}
\label{fig:fsig8}
\end{figure}

The \darklight 
contribution to this new research path has been 
presented recently \cite{hawken17}. By modelling the void-galaxy
cross-correlation function of VIPERS, a further complementary measurement of the 
growth rate of structure has been obtained\cite{hawken17}. This value  
is plotted in Fig.~\ref{fig:fsig8}, which provides a summary of 
all VIPERS estimates, plotted in the customary form
$f\sigma_8$ (see Sect.~\ref{sec:mod} for details). The figure also 
includes one further measurement, based on a joint analysis of RSD 
and galaxy-galaxy lensing \cite{delatorre17}, which has not been discussed here.
In addition, one more analysis is in progress, based on the 
  linearisation technique called ``clipping'' \cite{wilson16}.  
  
  Such a multifaceted approach to estimating the growth rate of structure clearly
  represents an important cross-check of residual systematic errors in each single 
  technique.  We stress again how this has been made possible thanks to the 
  broad ``information content'' of the VIPERS survey, which provides us with an optimal
  compromise (for these redshifts) between a large volume, a
  high sampling rate and extensive information on galaxy physical properties. 

%

\section{Optimal methods to derive cosmological parameters} 
\label{sec:bayes}
%
The cosmological information we are interested in is encoded in the two-point
statistics of the matter density field, i.e. its correlation function or, in Fourier space, 
its power spectrum $P_{\delta\delta}$. As we have seen in the Introduction,
this is connected to the observed galaxy fluctuations as $P_{gg}(k)=b^2P_{\delta\delta(k)}$, with
$n_g = \bar{n} \left(1 + b \delta \right)$.  The galaxy bias $b$ depends in general on the galaxy
properties, such as their luminosity and morphology, as well as
the environment in which they are found (in groups or in isolation).
Thus, in this context the bias terms are nuisance parameters that are
marginalized in the analysis.  However, the precision with which the
measurement can be made depends very much on these parameters as they set the
amplitude of the power spectrum and the effective signal-to-noise ratio.

Going beyond the standard approach to estimate cosmological parameters, as e.g. used in the $P(k)$ analysis of Fig.~\ref{fig:pk}, in \darklight we have investigated and applied optimal methods 
given the observed constraints (luminosity function and bias).  We can formulate
this as a forward modelling problem through Bayes' theorem, which tells us how
the measurements relate to the model:
\begin{equation}
p\left(P_{\delta\delta},\delta,b,\bar{n}|n_g\right) \propto p\left(n_g|P_{\delta\delta},\delta,b,\bar{n}\right) p\left(P_{\delta\delta},\delta,b,\bar{n}\right).
\end{equation}
On the left-hand side, the posterior describes the joint distributions of the
model parameters, here explicitly written as the density field ${\bf
\delta}$, its power spectrum $P_{\delta\delta}$, the galaxy bias ${\it b}$ and the
mean number density ${\it \bar{n}}$, but we can generalize to the underlying
cosmological parameters.  The posterior is factored into the likelihood and
prior terms on the right-hand side.  To evaluate the posterior we must assume
forms for these functions.  We begin by assuming multi-variate Gaussian
distributions for the likelihood and priors since these forms fully encode the
information contained in the power spectrum or correlation function
statistics.  In this limit the maximum-likelihood solution is given by the
Wiener filter.  In \cite{granett15} we demonstrate that in this limit the
solution is optimal in the sense that it minimizes the variance on the density
field and power spectrum.
\begin{figure}
\vspace{-0.4cm}
\centering
\includegraphics[scale=0.6]{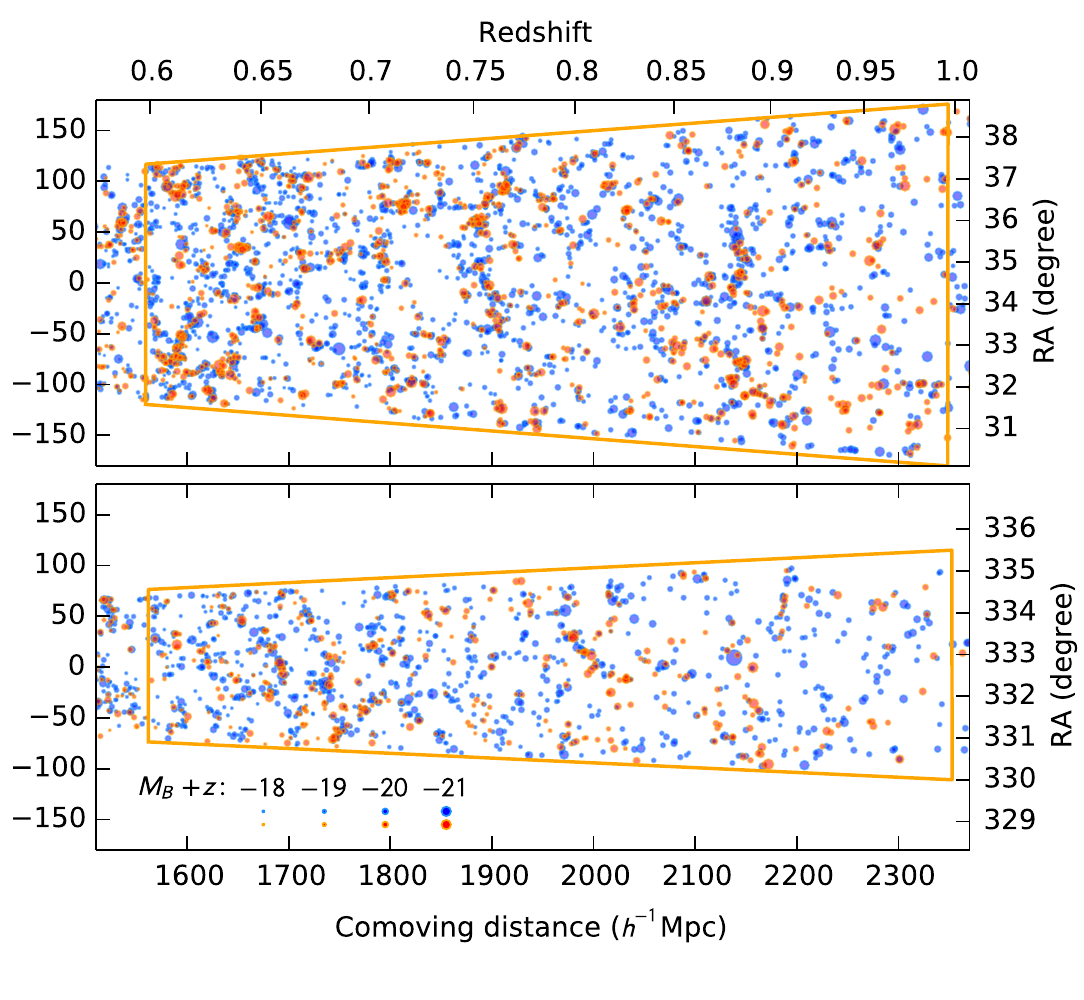}\includegraphics[scale=0.6]{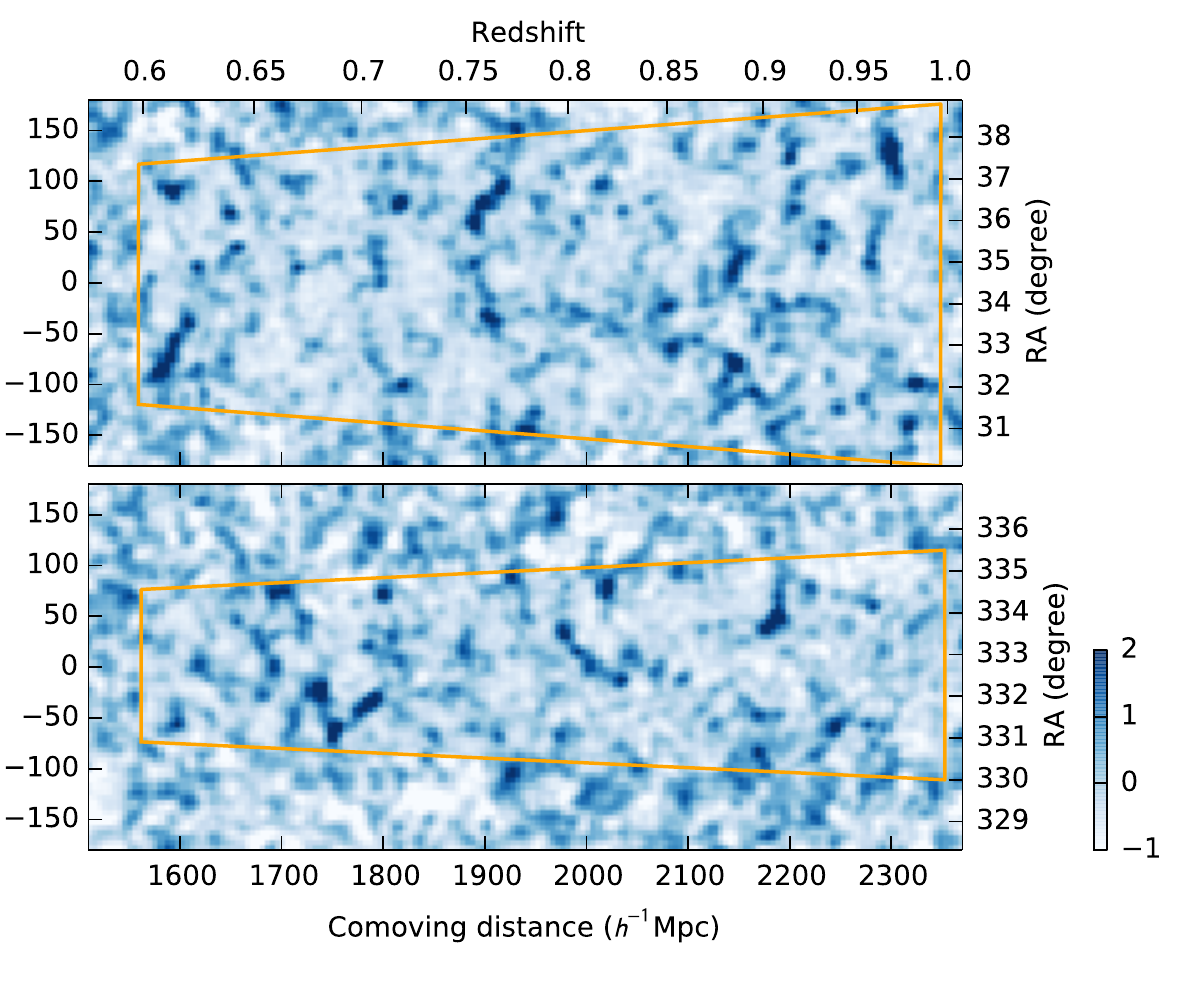}
\caption{{\it Left}: each point in these slices through the VIPERS W1 (top) and W4 (bottom) fields
represents a galaxy with a particular luminosity and color.  {\it Right}: the Wiener reconstruction
of the density field computed from these observations.  Beyond the survey limits the density
field is generated with a constrained random Gaussian field.
See \cite{granett15} for details.}
\label{fig:dens}
\end{figure}

Fig. \ref{fig:dens} shows one possible reconstruction of the VIPERS density
field. It represents a single step in the Monte Carlo chain used to sample the
full posterior distribution as presented in \cite{granett15}.  In this
work we characterized the full joint posterior likelihood of  the density
field, the matter power spectrum, RSD parameters, linear bias and luminosity
function.  These terms, particularly since they are estimated from a single
set of observations, are correlated and the analysis naturally reveals these
correlations.

A notable aspect of this analysis is that we optimally use diverse information
including the luminosity function, density field and power spectrum to infer
cosmological parameters and it becomes even more interesting with
additional observables.  We can envision simultaneous inference using cluster
counts or cosmic shear.  Generalizing requires putting a full dynamical model
for large-scale structure in the likelihood term effectively moving the 
likelihood analysis to the initial conditions. Observational systematics 
may be naturally included as well.

\section{A new kid in town: massive neutrinos}
\label{sec:neutrinos}
The non-vanishing neutrino mass, implied by the discovery of neutrino flavour
oscillations, has important consequences for our analysis of the
large-scale structure in the Universe.  Even if sub-dominant, the neutrino
contribution suppresses to some extent the growth of fluctuations on specific
scales, producing a deformation of the shape of the total matter power spectrum.
Given current upper limits on the sum of the masses $M_\nu$ ($M_\nu\leq 0.16$ eV 
at $95$\% confidence \cite{alam17}), the
expected effect corresponds to a few percent change in the amplitude
of total matter clustering. In the era of precision cosmology, neutrinos 
are an ingredient that cannot be neglected anymore. Conversely, future
surveys like Euclid may eventually be able to obtain an estimate of the total
mass of neutrinos with a precision that surpasses ground-based
experiments \cite{carbone11}.  To achieve this goal, we shall be able to: (a) describe how these
effects are mapped from the matter to the galaxy power spectrum,
i.e. what we measure; (b) distinguish these spectral deviations from
those due to non-linear clustering, and to the presence of other possible 
contributions, e.g. forms of dark energy beyond the cosmological
constant, like quintessence or in general an evolving equation of state of dark energy $w(z)$.
\begin{figure}
\centering
\includegraphics[scale=0.3]{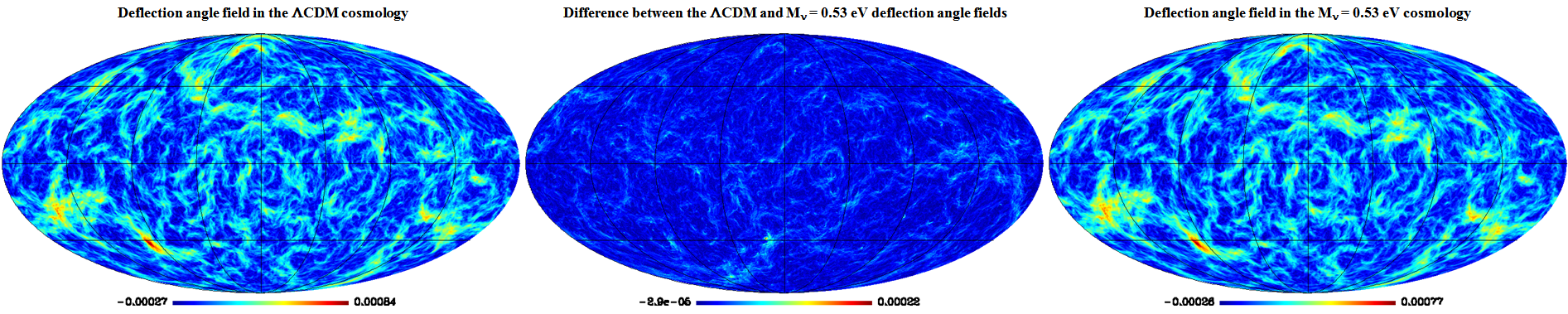}
\caption{Synthetic weak-lensing maps obtained via ray-tracing across the matter distribution in the DEMNUni simulations, for lensed sources spherically placed at $z=1$ around the observer. The left and right maps show the amplitude of the deflection angle for pure $\Lambda$CDM and for the case with $M_\nu=0.53$ eV total neutrino mass, respectively. The difference of the two maps is shown in the middle, and represents a change of about 6\% both in the {\it mean} and {\it rms} values of the deflection angle fields between the two scenarios.} 
\label{demnuni_maps}
\end{figure}

This has been addressed in \darklight through the ''Dark Energy and Massive Neutrino Universe'' (DEMNUni) simulations, a suite of fourteen large-sized N-body runs including massive neutrinos (besides cold dark matter), which have been recently completed \cite{carbone16}. They explore the impact on the evolution of structure of a neutrino component with three different total
masses ($M_\nu=0.17, 0.30, 0.53$ eV), including scenarios with evolving $w(z)$, according to the phenomenological form $w(z)= w_0 + w_a  z/(1+z) $.  

Running these simulations required developing new techniques to account for the evolving hot dark matter component represented by neutrinos \cite{zennaro17}. Early analyses of the whole suite show that the effects of massive neutrinos and evolving dark energy are highly degenerate (less than $2$\% difference) with a pure $\Lambda$CDM model, when one considers the clustering of galaxies or weak lensing observations.  Disentangling these different effects will therefore represent a challenge for future galaxy surveys as Euclid and needs to be carefully addressed.

Fig.~\ref{demnuni_maps} gives an example of physical effects that can be explored using these numerical experiments, showing weak-lensing maps (in terms of the amplitude of the resulting deflection angle) built via ray-tracing through the matter particle distribution of the simulations, for sources placed at redshift $z=1$.  The middle panel shows the difference between a pure $\Lambda$CDM scenario and a model with $M_\nu=0.53$ eV. More quantitatively, in terms of angular power spectra of the deflection field, massive neutrinos produce a scale-dependent suppression with respect to the $\Lambda$CDM case,  which, on small scales, asymptotically tends towards a constant value of about $10$\%, $19$\%, $31$\% for $M_\nu = 0.17, 0.30, 0.53$ eV, respectively. 

\paragraph{Acknowledgments.}
Many of the results presented here would have not been possible without the outstanding effort of the VIPERS team to  build such a unique galaxy sample. We are particularly grateful to S. de la Torre and J.A. Peacock for their insight and crucial contribution to the cosmological analyses discussed in this paper. Scientific discussions and general support in the development of Darklight by J. Dossett, J. He, and J. Koda are also warmly acknowledged.

\bibliographystyle{physics}
\bibliography{guzzo_etal}

\begin{thebibliography}{10}
\newcommand{\enquote}[1]{``#1''}

\bibitem{eisenstein11}
{Eisenstein}, D.~J., et~al.
\newblock \aj 142, 72 (2011).

\bibitem{guzzo_messenger17}
{Guzzo}, L., {Vipers Team}.
\newblock The Messenger 168, 40 (2017).

\bibitem{guzzo14}
{Guzzo}, L., et~al.
\newblock \aap 566, A108 (2014).

\bibitem{garilli14}
{Garilli}, B., et~al.
\newblock \aap 562, A23 (2014).

\bibitem{scodeggio17}
{Scodeggio}, M., et~al.
\newblock \aap, in press, ArXiv e-print 161107048  (2017).

\bibitem{york00}
{York}, D.~G., et~al.
\newblock \aj 120, 1579 (2000).

\bibitem{eisenstein05}
{Eisenstein}, D.~J., et~al.
\newblock \apj 633, 560 (2005).

\bibitem{diporto16}
{Di Porto}, C., et~al.
\newblock \aap 594, A62 (2016).

\bibitem{rota17}
{Rota}, S., et~al.
\newblock \aap 601, A144 (2017).

\bibitem{planck15}
{Planck Collaboration}, et~al.
\newblock ArXiv e-print: 150201589  (2015).

\bibitem{riess98}
{Riess}, A.~G., et~al.
\newblock \aj 116, 1009 (1998).

\bibitem{perlmutter99}
{Perlmutter}, S., et~al.
\newblock \apj 517, 565 (1999).

\bibitem{cole05}
{Cole}, S., et~al.
\newblock \mnras 362, 505 (2005).

\bibitem{alam17}
{Alam}, S., et~al.
\newblock \mnras 470, 2617 (2017).

\bibitem{davis83}
{Davis}, M., {Peebles}, P.~J.~E.
\newblock \apj 267, 465 (1983).

\bibitem{kaiser87}
{Kaiser}, N.
\newblock \mnras 227, 1 (1987).

\bibitem{guzzo08}
{Guzzo}, L., et~al.
\newblock \nat 451, 541 (2008).

\bibitem{laureijs11}
{Laureijs}, R., et~al.
\newblock ArXiv e-print 11103193  (2011).

\bibitem{okumura11}
{Okumura}, T., {Jing}, Y.~P.
\newblock \apj 726, 5 (2011).

\bibitem{bianchi12}
{Bianchi}, D., et~al.
\newblock \mnras 427, 2420 (2012).

\bibitem{peacock01}
{Peacock}, J.~A., et~al.
\newblock \nat 410, 169 (2001).

\bibitem{peacock94}
{Peacock}, J.~A., {Dodds}, S.~J.
\newblock \mnras 267, 1020 (1994).

\bibitem{pezzotta17}
{Pezzotta}, A., et~al.
\newblock \aap 604, A33 (2017).

\bibitem{delatorre12}
{de la Torre}, S., {Guzzo}, L.
\newblock \mnras 427, 327 (2012).

\bibitem{taruya10}
{Taruya}, A., {Nishimichi}, T., {Saito}, S.
\newblock \prd 82, 6, 063522 (2010).

\bibitem{bianchi15}
{Bianchi}, D., {Chiesa}, M., {Guzzo}, L.
\newblock \mnras 446, 75 (2015).

\bibitem{fisher95}
{Fisher}, K.~B.
\newblock \apj 448, 494 (1995).

\bibitem{scoccimarro04}
{Scoccimarro}, R.
\newblock \prd 70, 8, 083007 (2004).

\bibitem{reid12}
{Reid}, B.~A., et~al.
\newblock \mnras 426, 2719 (2012).

\bibitem{bianchi16}
{Bianchi}, D., {Percival}, W.~J., {Bel}, J.
\newblock \mnras 463, 3783 (2016).

\bibitem{jennings11}
{Jennings}, E., {Baugh}, C.~M., {Pascoli}, S.
\newblock \mnras 410, 2081 (2011).

\bibitem{bel17}
{Bel}, J., et~al.
\newblock \rm{in preparation}  (2017).

\bibitem{blake10}
{Blake}, C., et~al.
\newblock \mnras 406, 803 (2010).

\bibitem{mohammad17}
{Mohammad}, F.~G., et~al.
\newblock ArXiv e-prints  (2017).

\bibitem{hamaus16}
{Hamaus}, N., et~al.
\newblock Physical Review Letters 117, 9, 091302 (2016).

\bibitem{micheletti14}
{Micheletti}, D., et~al.
\newblock \aap 570, A106 (2014).

\bibitem{hawken17}
{Hawken}, A.~J., et~al.
\newblock \aap, in press, ArXiv e-print 161107046  (2017).

\bibitem{delatorre17}
{de la Torre}, S., et~al.
\newblock submitted to \aap, ArXiv e-prints 161205647  (2017).

\bibitem{wilson16}
{Wilson}, M., et~al.
\newblock \rm{in preparation}  (2017).

\bibitem{blake12}
{Blake}, C., et~al.
\newblock \mnras 425, 405 (2012).

\bibitem{beutler17}
{Beutler}, F., et~al.
\newblock \mnras 466, 2242 (2017).

\bibitem{beutler12}
{Beutler}, F., et~al.
\newblock \mnras 423, 3430 (2012).

\bibitem{howlett15}
{Howlett}, C., et~al.
\newblock \mnras 449, 848 (2015).

\bibitem{okumura16}
{Okumura}, T., et~al.
\newblock Pub Astr Soc Japan 68, 38 (2016).

\bibitem{granett15}
{Granett}, B.~R., et~al.
\newblock \aap 583, A61 (2015).

\bibitem{carbone11}
{Carbone}, C., et~al.
\newblock \jcap 3, 030 (2011).

\bibitem{carbone16}
{Carbone}, C., {Petkova}, M., {Dolag}, K.
\newblock \jcap 7, 034 (2016).

\bibitem{zennaro17}
{Zennaro}, M., et~al.
\newblock \mnras 466, 3244 (2017).

\end{thebibliography}

\end{document}